\documentclass[a4paper,onecolumn,11pt,accepted=2025-05-27]{quantumarticle}
\pdfoutput=1

\usepackage{amsfonts}
\usepackage{amsmath}
\usepackage{amssymb}
\usepackage{amsthm}

\usepackage{hyperref}
\usepackage{gensymb}
\usepackage{xcolor}
\usepackage{braket}
\usepackage{graphicx}
\usepackage{units}

\hypersetup{colorlinks=true}
\hypersetup{citecolor=blue}
\hypersetup{urlcolor=blue}

\newcommand{\omitted}{{\rule{1ex}{.4pt}}}
\DeclareMathOperator\Prob{Prob}
\DeclareMathOperator\erfc{erfc}
\newcommand*\ketbra[2]{\ket{#1}\!\!\bra{#2}}

\usepackage{comment}
\usepackage[normalem]{ulem}

\begin{document}


\title{How to avoid (apparent) signaling in Bell tests}


\author{Massimiliano Smania}
\email{massimiliano.smania@gmail.com}
\affiliation{Department of Physics, Stockholm University, S-10691 Stockholm, Sweden}

\author{Matthias Kleinmann}
\email{matthias.kleinmann@uni-siegen.de}
\affiliation{Naturwissenschaftlich-Technische Fakult\"at, Universit\"at Siegen, Walter-Flex-Stra{\ss}e 3, D-57068 Siegen, Germany}

\author{Ad\'an Cabello}
\email{adan@us.es}
\affiliation{Departamento de F\'{\i}sica Aplicada II, Universidad de Sevilla, E-41012 Sevilla, Spain}
\affiliation{Instituto Carlos~I de F\'{\i}sica Te\'orica y Computacional, Universidad de Sevilla, E-41012 Sevilla, Spain}

\author{Mohamed Bourennane}
\email{boure@fysik.su.se}
\affiliation{Department of Physics, Stockholm University, S-10691 Stockholm, Sweden}


\begin{abstract}
Bell tests have become a powerful tool for quantifying security, randomness, entanglement, and many other properties, as well as for investigating fundamental physical limits. In all these cases, the specific experimental value of the Bell parameter is important as it leads to a quantitative conclusion. However, experimental implementations can also produce experimental data with (apparent) signaling. This signaling can be attributed to systematic errors occurring due to weaknesses in the experimental designs. Here we point out the importance, for quantitative applications, to identify and address this problem. We present a set of experiments with polarization-entangled photons in which we identify common sources of systematic errors and demonstrate approaches to avoid them. In addition, we establish the highest experimental value for the Bell-CHSH parameter obtained after applying strategies to minimize signaling that we are aware of: $S = 2.812 \pm 0.003$ and negligible systematic errors. The experiments did not randomize the settings and did not close the locality loophole.
\end{abstract}

\maketitle


\section{Motivation}


Bell inequalities were conceived to test whether quantum theory can be replaced by a local hidden-variable theory \cite{Bell64}. However, today the role and purpose of Bell tests has changed and the particular value of the Bell parameter is used to quantify, for example, the rate of the device-independent secure key generation \cite{ABGMPS07, ARV16, NDN22,ZVR22}, the degree of entanglement \cite{GT09, MBLHG13}, the advantages in communication complexity \cite{BZPZ04, BCMW10}, the robustness of incompatibility \cite{HKR15, CBLC16, CS16}, the dimension \cite{BPAGMS07}, the degree of self-testing \cite{PR92, BP15, K16}, and the number of private random bits \cite{AM16}. The experimental value is also crucial when we investigate the physical limits of correlations \cite{CMACGLMSZNBLGK13, CLBGK15, PJC15} and test how close we can get to the upper bound according to quantum theory \cite{Tsirelson80} or whether it is even possible to exceed this bound.

Bell tests are based on the assumption of nonsignaling, that is, that two independent, separated experiments cannot influence each other. This assumption is so natural that there seems to be little value in making the effort of testing whether it is satisfied. However, as we demonstrate here, various different systematic errors in experiments can be held responsible for producing apparent signaling. 
The apparent signaling considered here is statistically significant and goes beyond the nonsignaling conditions arising from statistical fluctuations. It should be noted that our work concerns Bell experiments where fair--sampling is an assumption. In contrast, in the loophole--free Bell tests \cite{HBD15, GVW15, SMC15, W16}, all known effects that could create apparent signaling have been eliminated and consequently the data is free of this defect \cite{HBD15, GVW15, SMC15, W16}.

The aim of this article is to identify common sources of systematic errors which can distort the value of the Bell parameter or cause apparent signaling, and describe how to avoid them. Our list cannot be comprehensive, but it is representative of typical errors. Our aim is to point out that for all applications where a precise estimation of the Bell parameter is required and fair--sampling is one of the assumptions, a deeper analysis of apparent signaling, and the elimination of it, are crucial in order to have reliable results. Some systematic errors can be compensated for by post-processing, but doing so in a consistent and reliable manner needs a careful analysis which may be only acceptable in case of small residual errors. As will be shown in the following, this is far from being the case in usual experimental implementations. Although we focus on photonic implementations and the Clauser--Horne--Shimony--Holt (CHSH) scenario \cite{CHSH69}, which is maybe the most common combination for practical applications, our analysis and conclusions also apply to other Bell-like scenarios.


\section{The CHSH scenario and the nonsignaling conditions}


We consider two independent experimenters, Alice and Bob, each of them able to perform measurements $x=0,1$ for Alice, and $y=0,1$ for Bob. All measurements have two outcomes $a =\pm 1$ for Alice, and $b =\pm 1$ for Bob. In the CHSH scenario, the Bell parameter is given by
\begin{equation}
S = E(0,0)-E(0,1)+E(1,0)+E(1,1),
\end{equation}
with $E(x,y)=\sum_{a,b}ab P(a,b|x,y)$ and $P(a,b|x,y)$ denoting the joint probability of Alice and Bob obtaining outcomes $a$ and $b$, respectively, when they measure $x$ and $y$, respectively. According to quantum theory $\lvert S\rvert\le 2\sqrt{2}$ holds as a universal bound \cite{Tsirelson80} and this bound can be saturated already for two-level systems, if both parties share a maximally entangled quantum state, such as $\ket{\Phi^+}= (\ket{00}+\ket{11})/\sqrt 2$.

For the CHSH scenario, the assumption that both parties measure independently implies the nonsignaling conditions
\begin{equation}\begin{split}
\alpha^A_{a,x}\equiv P(a,\omitted|x,0)-P(a,\omitted|x,1)=0
&\quad \text{for all $x,a$}\\
\alpha^B_{b,y}\equiv P(\omitted,b|0,y)-P(\omitted,b|1,y)= 0
&\quad \text{for all $y,b$.}
\end{split}\end{equation}
Here we wrote $P(a,\omitted|x,y)$ for $\sum_b P(a,b|x,y)$ and similarly for $P(\omitted,b|x,y)$. Due to the normalization of the correlations, $P(\omitted,\omitted|x,y)=1$ for all $x,y$, it is sufficient to consider the nonsignaling conditions only for $a=+1$ and 
$b=+1$, respectively.

In a straightforward test of experimental data, one considers directly these four independent nonsignaling conditions. For example, $\hat \alpha^A_{+1,0}$ is obtained by replacing the correlations in $\alpha^A_{+1,0}$ by the empirical frequencies. Even in an ideal experiment, the condition $\hat \alpha^A_{+1,0}=0$ will not be satisfied exactly, but rather fluctuate with a standard deviation of $\sigma^A_{+1,0}$ about $0$. However, if there is a systematic violation of a nonsignaling condition, then the ratio $\hat \alpha^A_{+1,0}/\sigma^A_{+1,0}$ will raise roughly with the square root of the number of total events, $\sqrt{n}$. This effect can be clearly seen, for example, in the right panel of Fig.~\ref{fig:results}~a).

For a joint test of the four nonsignaling conditions, comprehensive statistical methods are available, such as those suggested in Refs.~\cite{Moroder13, Liang19} for testing systematic errors and nonsignaling conditions. Here we use the likelihood ratio test: if the data was sampled from correlations $P(a,b|x,y)$ which obey the nonsignaling constraints, then the space $\mathcal X$ spanned by those correlations has reduced dimension compared to the space $\mathcal X_0$ which contains all correlations. Assuming Poissonian statistics and independent and identically distributed (i.i.d.) samples, for given data, we compute the maximum likelihood $\mathcal L$ for correlations restricted to $\mathcal X$ and similarly $\mathcal L_0$ for $\mathcal X_0$. The characteristic quantity $\xi= -2(\log \mathcal L-\log\mathcal L_0)$ is (as a very good approximation) $\chi_\delta^2$ distributed \cite{Wilks62} with the number of degrees $\delta$ being the difference of the dimensions of $\mathcal X_0$ and $\mathcal X$. We compute the $p$-value via $p=\Prob( t\ge \xi )$ for $\chi^2_\delta$ distributed $t$. Note, that a failure of the nonsignaling constraints would result in a larger value of $\xi$ and, then, in a lower $p$-value. Since those $p$-values can be very small, we transform them to equivalent standard deviations $z=\sqrt 2\erfc^{-1}(p)$, where $\erfc^{-1}(p)$ is the inverse of the complementary error function. In other words, we obtain $z$ equivalent standard deviations when the observed data is as unlikely as if, in an analogous experiment with normal distributed data, an outcome would have been observed with $z$ standard deviations away from the expected mean.


\begin{figure}\begin{minipage}{.495\linewidth}\centering
\includegraphics[width=.9\linewidth]{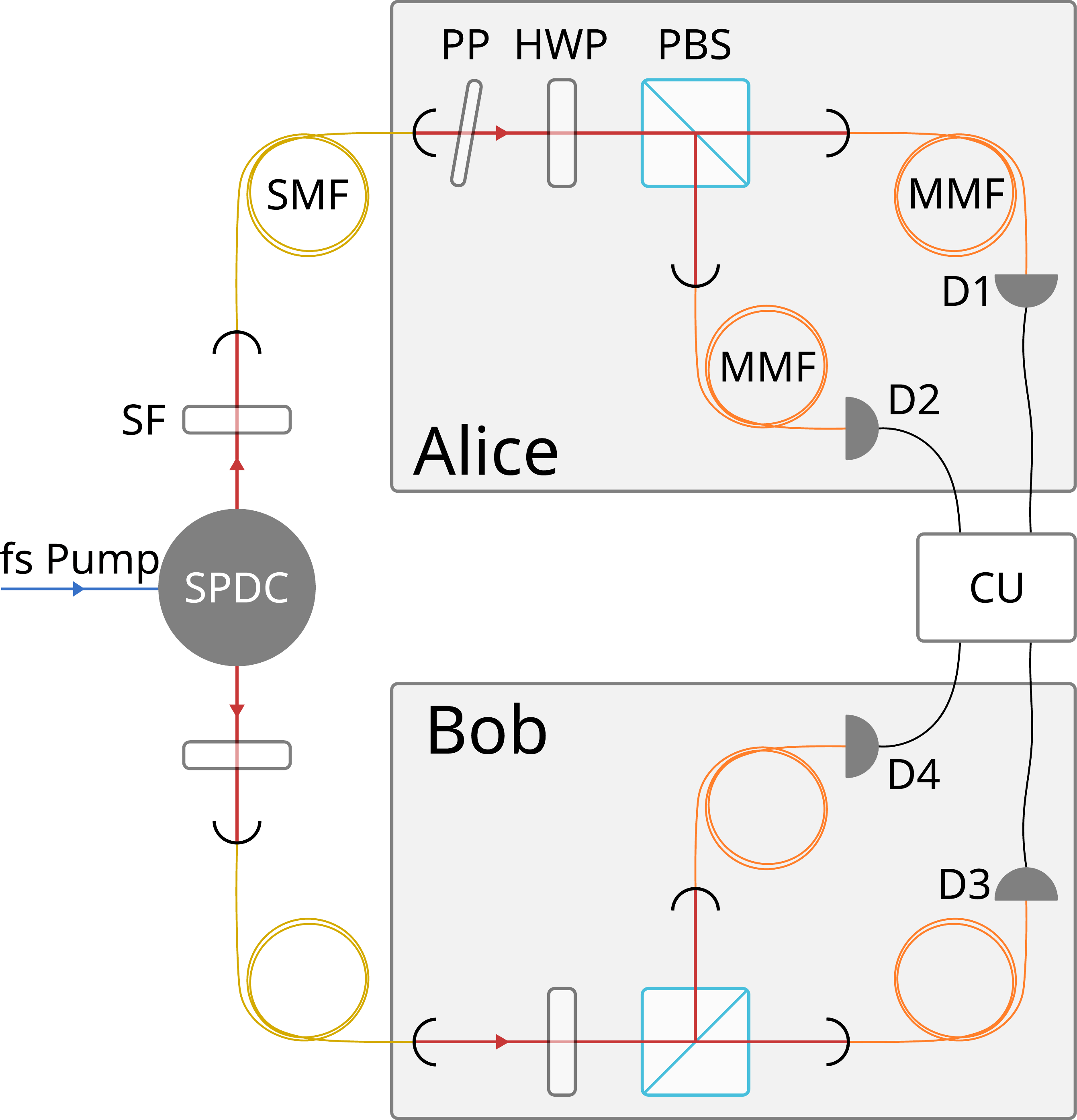}
\end{minipage}\begin{minipage}{.495\linewidth}
\caption{\label{fig:setup}%
Femtosecond laser pulses at \unit[390]{nm} are converted into pairs of polarization-entangled photons at \unit[780]{nm} through type--I spontaneous parametric down-conversion (SPDC) in two orthogonally oriented \unit[2]{mm} thick beta-barium borate crystals. After the SPDC, each of the two photons passes through \unit[1]{nm} narrow spectral filters (SF) and single-mode fibers (SMF). Alice's and Bob's measurement stations each consist of a half-wave plate (HWP), a polarizing beam splitter (PBS), multi-mode fibers (MMF), and single photon avalanche photodiodes (D1--D4). Alice's station also includes a phase plate (PP). Photon detection events are registered by a coincidence unit (CU). Rates of around 200 total coincidences per second were used. Through careful alignment, the Bell state $\ket{\Phi^+} =(\ket{00}+\ket{11})/\sqrt 2$ is prepared with a visibility of $0.994 \pm 0.001$ in the interference basis, throughout the experiments. The measurements on Alice's stations are $\sigma_z=\ketbra00-\ketbra11$ for setting $x=0$ and $\sigma_x=\ketbra01+\ketbra10$ for setting $x=1$; on Bob's station they are $(\sigma_z+\sigma_x)/\sqrt2$ for $y=0$ and $(\sigma_z-\sigma_x)/\sqrt2$ for $y=1$.}
\end{minipage}
\end{figure}


\section{Experimental sources of apparent signaling}


Here we identify experimental sources of apparent signaling and address them by adding components to the initial setup and optimizing the data acquisition procedure. We start by considering a common configuration for Bell experiments using polarization-entangled photon pairs generated through a down-conversion process. It consists of a free-space down-conversion source and two measurement stations, each including a polarizer and one single-mode fiber (SMF) connected to a single-photon detector. This is the specific configuration in which the highest values for $S$ to date have been achieved \cite{CMACGLMSZNBLGK13, CLBGK15, PJC15}. Nevertheless, it suffers from two key limitations that could lead to apparent signaling in the results:

(L1) Drifts in the pump laser power. The measurements are performed with only one detector on each side and the outcomes $\pm1$ are measured by choosing orthogonal settings for the polarizer. The data for the different measurement outcomes is obtained from different experimental sequences. If, during this process, the intensity of the pump laser drifts, then the empirical frequencies of the outcomes do not yield the correlations since the fair sampling assumption is not satisfied.

(L2) Polarization--dependent collection efficiency. Any optical element can slightly displace the beam if moved or rotated during the measurement. This is particularly true for wave plates or polarizers used to change the measurement settings. Careful alignment and particularly good components may reduce the problem, but if SMFs are used to collect the photons, this issue will still affect the coincidence rates depending on the measurement setting. The difficulty here is that SMFs have a core diameter of only a few micrometers, while multi-mode fibers (MMF) at the measurement stations would collect photons with different spatial modes and thus decrease the visibility.

A clean and straightforward solution for limitation (L1) is to use a polarizing beam splitter and two detectors on each measurement station. Then the normalization is naturally identical for all outcomes, independent of the laser 
intensity. Regarding limitation (L2), a practical solution consists in having SMFs between the source and the measurement stations, since here no component is moved during the measurement. MMFs are then used for collecting photons for the detection. The SMF acts as spatial mode filter, while the MMFs are effectively insensitive to small beam displacements due to their core being at least ten times wider. Such a setup, depicted in Fig.~\ref{fig:setup}, is our starting point, since it naturally neither suffers (L1) nor (L2).


\begin{figure}\centering
\includegraphics[width=.9\linewidth]{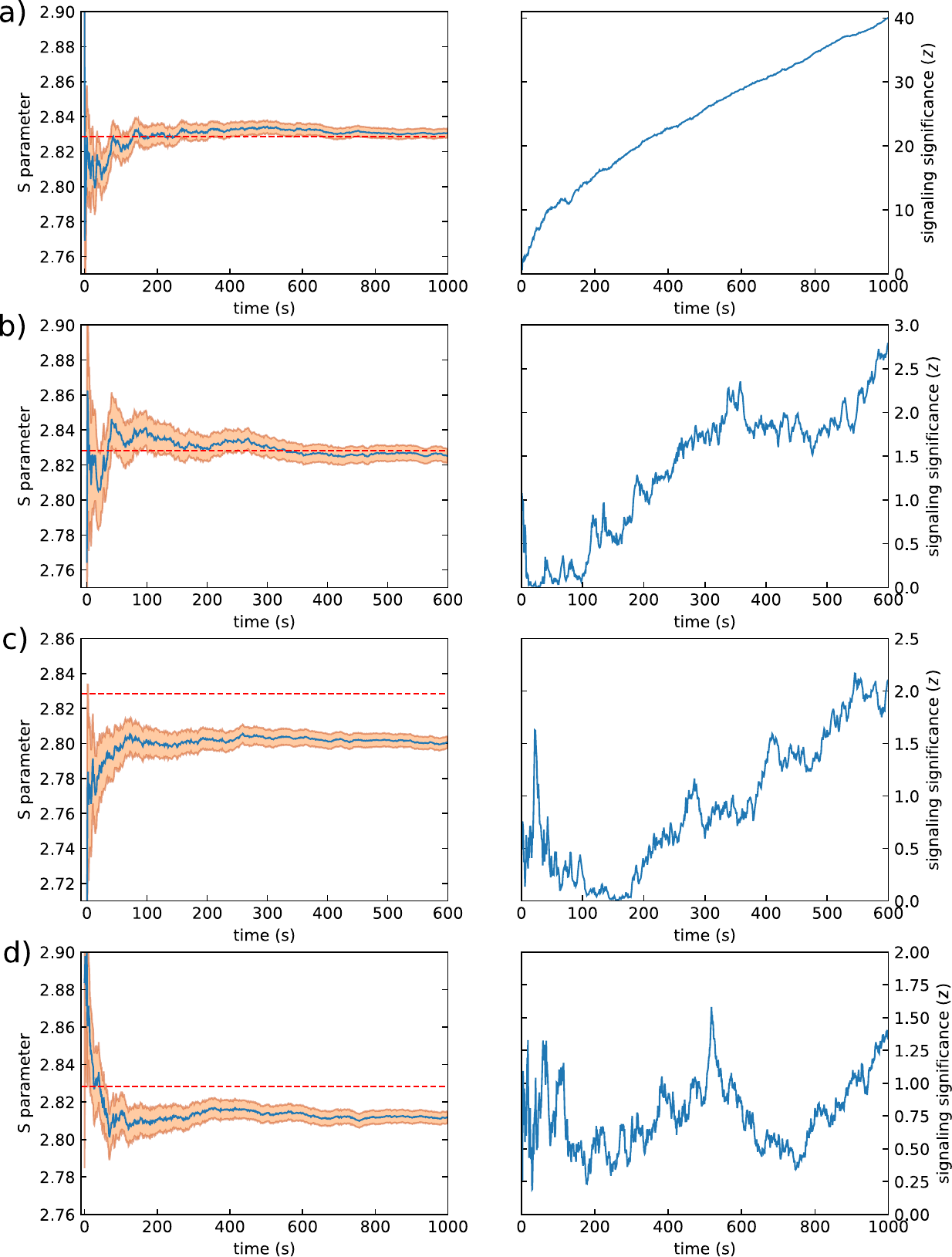}
\caption{\label{fig:results}%
Effect of different measures taken to address the systematic errors, as function of the data collection time. Left panels: Bell-CHSH parameter $S$ (solid blue line) with the corresponding statistical uncertainty (orange shaded area, one standard deviation) and the quantum maximum $S=2\sqrt2$ (dashed red line). Right panels: Statistical significance of the apparent signaling in equivalent standard deviations $z$. a) Simple setup with low setting reproducibility and asymmetric collection efficiency. b) Improved setup addressing setting reproducibility by using precise motors and repeating the settings 10 times. c) Improved setup addressing asymmetric collection efficiency and also using precise motors, but not repeating the settings. d) Setup used to obtain our experimental value, using precise motors, symmetric collection efficiency, and repeating the settings 200 times. See Table~\ref{tab:summary} for additional information.}
\end{figure}


In the first experiment we proceed as follows: First we maximize the visibility and the collection efficiency. After that, every setting is continuously measured for 1000 seconds. The order of the settings is $(x,y)=(0,0)$, $(0,1)$, $(1,1)$, and then $(1,0)$. The results of this first experiment are presented in Fig.~\ref{fig:results}~a). While the obtained value for $S$ is very high---within one sigma from the quantum maximum---, the statistical significance of the apparent signaling grows to be about $40$ standard deviations. For comparison, as there is a large number of experimental results involving the Bell parameter, in this work, as an example, we analyze the experiment aiming at the highest violation. In the analysis of the data of Ref.~\cite{PJC15}, we find that the nonsignaling constraints are violated by more than $200$~standard deviations. This result implies that the experiment suffered systematic effects that had not been taken into account and hence the error bars reported in Ref.~\cite{PJC15} are not supported by the data. The main reasons behind our result presented in Fig.~\ref{fig:results} a) are the following:

(S1) Measurement setting precision and measurement setting reproducibility. While a low accuracy will result in lower visibility and hence in a lower value of $S$, low precision in repeating a measurement setting also yields apparent 
signaling, as this is equivalent to Bob setting his polarizer differently according to Alice's measurement setting. In our first experiment, motors with a precision of $0.2\degree$ were used in order to simulate manual rotation of the wave plates.

(S2) Asymmetric collection efficiency. This may violate the fair sampling assumption. If the photons are collected with different efficiency at the two outputs of the same measurement station, signaling may appear in the data. To get some intuition, suppose that the ``$-1$'' detector of Alice has efficiency $\eta<1$ while all other detectors have unit efficiency. For the state $\ket{\Phi^+}$, Alice's setting $x=0$ ($x=1$) corresponding to a measurement of $\sigma_z$ ($\sigma_x$), respectively, and Bob's $y=1$ setting to $(\sigma_z-\sigma_x)/\sqrt2$, one finds $\alpha^B_{+1,y}=\sqrt{1/2}(1-\eta)/(1+\eta)\ne 0$. In our first experiment the overall collection efficiency was maximized, resulting in a different efficiency for each detector.

The second experiment aims to address (S1). An obvious countermeasure to the reproducibility issue is to use more precise motors. Therefore, in the second experiment we used motors with a precision of $0.02\degree$. Nevertheless, no matter how precise those motors are, they will always stop slightly short or long of the desired position, therefore yielding, for large number of samples, a violation of the nonsignaling conditions. A further approach to circumvent (S1) is to measure each setting with a short collection time and to repeat the experiment many times. If we assume that the motor precision is sufficiently symmetric around a central position, then the deviations from the desired angle will average out. Therefore, in the second experiment, each sequence of the four settings was repeated 10 times. The results of the second experiment are presented in Fig.~\ref{fig:results}~b), and show that the improvements have been effective in reducing the apparent signaling. Still, the unequal collection efficiency causes the significance of the signaling to grow as the collection time 
increases.


\begin{figure}\centering
\includegraphics[width=\linewidth]{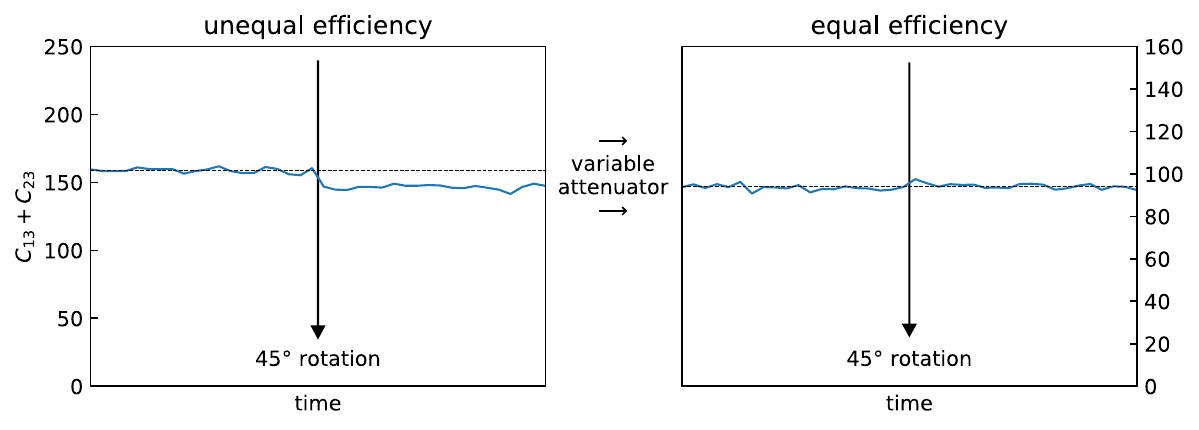}
\caption{\label{fig:case2deteff}%
Left: Different total coincidence rates on Alice's side indicate differences in the detection efficiencies of D1 and D2. Right: A variable attenuator in front of the detector with higher efficiency compensates this defect.}
\end{figure}


In order to investigate the effect of asymmetric collection efficiency (S2), we perform a third experiment in which we compensate the different efficiencies for each party (D1 vs.\ D2 and D3 vs.\ D4) by using variable attenuators in front of the MMF couplers. While this issue can in principle be addressed in the data analysis through additional post--processing, we chose to take care of the problem at the source, in the experimental setup. To achieve this, we define as $C_{ij}(\theta)$ the rate of coincident clicks of detectors D$i$ and D$j$ when Alice's wave plate is at angle $\theta$. Then, we compare the rates 
$\gamma=C_{13}(0\degree)+C_{23}(0\degree)$ and 
$\gamma'=C_{13}(45\degree)+C_{23}(45\degree)$, see Fig.~\ref{fig:case2deteff}. 
If $\gamma \ne \gamma'$, then D1 and D2 have different detection rates. In this case, we use a variable attenuator in the path with higher detection rate in order to approach $\gamma = \gamma'$. In our third experiment, we implement this procedure for Alice and Bob; we use the more precise motors ($0.02\degree$), but we do not repeat the settings. The results are shown in Fig.~\ref{fig:results}~c) and show that compensating asymmetric collection efficiencies reduced the apparent signaling to $2.1$ standard deviations.

In the fourth experiment we improved the third experiment by also repeating each sequence of the four settings 200 times. The results of this fourth experiment are shown in Fig.~\ref{fig:results}~d). There, the statistical significance of the apparent signaling is $1.3$ standard deviations, which indicates that no major source of signaling is present for the amount of samples taken. The results of all four experiments are summarized in Table~\ref{tab:summary}.


\begin{table}\centering
\begin{tabular}{rlcllcrr}
&
signaling sources & $S$ & $\sigma_\text{stat}$ & $\sigma_\text{syst}$ & $z$ & $T_\mathrm{sett}$ & repetitions\\\hline
a) & P, E, R    & 2.8305 & 0.0026 & 0.02   & 40\phantom{.0}   & \unit[1000]{s} &   1\\
b) & E          & 2.8252 & 0.0041 & 0.005  & \phantom02.8 &   \unit[60]{s} &  10\\
c) & R          & 2.8006 & 0.0035 & 0.002  & \phantom02.1 &  \unit[600]{s} &   1\\
d) & none major & 2.8117 & 0.0032 & 0.0001 & \phantom01.3 &    \unit[5]{s} & 200\\
\end{tabular}
\caption{\label{tab:summary}%
Summary of the four experiments a)--d) in Fig.~\ref{fig:results}.
Signaling sources are: low motor precision ``P'', no repetition of the measurement settings ``R'', and asymmetric collection efficiency ``E''.
The value of the Bell-CHSH parameter $S$ is given with the statistical standard deviation $\sigma_\text{stat}$, as obtained from error propagation and assuming Poissonian distributions, and the systematic error $\sigma_\text{syst}$. The statistical significance $z$ of the apparent signaling is computed using a likelihood ratio test and is given as equivalent standard deviations.
The coincidence rate is around $200$ coincidences per second. The settings were measured in the order $x=0,y=0$ then $x=0,y=1$, then $x=1,y=0$, then $x=1,y=1$ and each setting was measured for the time $T_\mathrm{sett}$. This procedure was repeated according to the last column, yielding about $200\times 10^3$ coincidences per setting for experiments a) and d) and about $120\times 10^3$ for b) and c).}
\end{table}


\section{Estimate for the systematic errors}


In the analysis so far, we considered the statistical significance of the apparent signaling as an indicator for systematic errors dominating the experimental evaluation. However, conversely, when we do not observe any significant apparent signaling this does not purge us of a quantitative analysis of sources of systematic errors in the experiment, which may affect our estimate of the Bell-CHSH parameter $S$. In our experiments the number of samples was such that the statistical uncertainty originating from Poissonian shot noise was of the order of $0.003$. We compare this to the following sources of systematic errors:

(E1) Motor precision. As we demonstrated, the motor precision has a major contribution to the total systematic error. Whenever a setting has to be repeated (that is, a wave plate has to go back to a previous position after 
having been moved), an uncertainty in the wave plate angle affects the result of the experiment. When low precision motors were used, as in our first experiment, then the propagated uncertainty on the Bell-CHSH parameter $S$ due to 
their precision of $0.2\degree$ is $0.02$, which is larger than the statistical error. For the other experiments, motors with a precision of $0.02 \degree$ 
were used. This gives a propagated systematic error on $S$ of $0.002$ and is comparable to the statistical uncertainty. However, when the measurement settings are repeated $N$ times, this error is suppressed by a factor of $1/\sqrt N$.

(E2) Detector dark counts. Each avalanche photodiode used in the experiment has a dark count rate of approximately 500 clicks per second. The pump pulsed laser has a width of 400 to \unit[500]{fs} and a repetition rate of \unit[80]{MHz}. The coincidence count window is \unit[1.6]{ns}, and the single and coincidence count rates are 1000 and 200 per second, respectively. The rate of accidental coincidences results can then be estimated to be $10^{-5}$.

(E3) Higher order down-conversion events. At the low pump power used in the experiment, the rate of so-called accidental coincidences was fairly minimal, of the order of 0.1 per second or less. This can be estimated by counting 
coincident events between detectors in the same measurement station. Such a low rate does nevertheless affect the final result by lowering visibilities and therefore the violation. This can be seen as if the source would be more noisy. 
Therefore, we do not consider this as a systematic error and we do not subtract this noise in our evaluation.


\section{Conclusions}


Experimental setups aiming for high values of the Bell parameter are susceptible to systematic errors causing apparent signaling and, therefore, rendering a quantitative estimation of the Bell parameter questionable. Specifically, these systematic errors cannot be explained by shot noise, but rather appear due to common limitations in the experimental setups and typical methods for acquiring the experimental data. For modern quantitative 
applications of Bell tests, these problems have to be addressed and accounted for to obtain universally viable estimates for the Bell parameter. Here we have 
experimentally identified the most common sources of systematic errors leading to apparent signaling and described how to address (that is, minimize the 
impact of) them. We have obtained a value for the Bell-CHSH parameter of $S=2.812 \pm_\text{stat} 0.003$ and negligible systematic errors. This is the highest value we know that is obtained after applying strategies to minimize signaling. For future Bell experiments aimed at quantitative applications, we should avoid the sources of 
systematic errors described in this paper as much as possible and understand and estimate residual systematic errors.


\section{Acknowledgments}


We are grateful to M.\ Nawareg for discussions and for optimizing the 
entanglement source. We thank
A.\ Cer\`e,
B.G.\ Christensen,
N.\ Gisin,
S.K.\ Joshi,
C.\ Kurtsiefer,
Y.-C.\ Liang,
G.\ Lima,
M.W.\ Mitchell,
H.S.\ Poh,
R.\ Renner, and
G.\ T\'oth
for discussions and feedback, and to the authors of Ref. \cite{PJC15} for giving us the opportunity of analyzing their data.
This work was supported by the EU (project \href{https://doi.org/10.3030/101070558}{FoQaCiA}, ERC 
Starting Grant No.\ GEDENTQOPT and Consolidator Grant No.\ 683107/TempoQ), the 
\href{https://doi.org/10.13039/501100011033}{MCINN/AEI} (Project No.\ PID2020-113738GB-I00),
the DFG (Forschungsstipendium KL~2726/2-1), and the Wallenberg Center for Quantum Technology (WACQT).



\end{document}